\documentclass[12pt,english]{article}
\usepackage[LGR,T1]{fontenc}
\usepackage[latin9]{inputenc}
\usepackage{geometry}
\usepackage{amsmath}
\usepackage{amssymb}

\makeatletter

\DeclareRobustCommand{\greektext}{%
  \fontencoding{LGR}\selectfont\def\encodingdefault{LGR}}
\DeclareRobustCommand{\textgreek}[1]{\leavevmode{\greektext #1}}
\ProvideTextCommand{\~}{LGR}[1]{\char126#1}

\date{}

\makeatother

\usepackage{babel}
\begin{document}

\title{\textgreek{L}CDM cosmology from visible matter only}

\author{{\normalsize{}Herman Telkamp}\thanks{{\footnotesize{}Jan van Beverwijckstr. 104, 5017 JA Tilburg, Netherlands,
email: herman\_telkamp@hotmail.com}}}
\maketitle
\begin{abstract}
\noindent We discuss physical interpretation of \textgreek{L}CDM cosmology
from a Machian model of the universe containing nothing but visible
matter (ordinary matter, radiation). The Friedmann equation can be
derived from a Machian definition of energy, whereby both kinetic
and potential energy of a particle are related to all cosmic matter-energy
within the particle's gravitational horizon. The distance to this
horizon thus appears as a parameter in all forms of matter-energy
density. From conservation of Machian energy it follows that all different
types of matter-energy are uniformly characterized by $\rho\propto a^{-1}$,
i.e., by a constant deceleration $q=-1/2$. This coincides with relative
densities $\varOmega_{m}=1/3$ and $\Omega_{\Lambda}=2/3$ of \textgreek{L}CDM.
Thus the Machian cosmological model matches present relative densities
of \textgreek{L}CDM, without invoking dark components. \bigskip{}
\end{abstract}
The \textgreek{L}CDM cosmological model matches a wide range of observations
quite accurately. Despite this success, the model relies on the existence
of hypothesized dark matter and dark energy, so its physical justification
is poor. We therefore consider physical interpretation of the (flat)
\textgreek{L}CDM model with Friedmann equation 

\begin{equation}
\frac{\dot{a}^{2}}{a^{2}}=\frac{8}{3}\pi G(\rho+\rho_{\Lambda}),\label{eq:Fr LCDM}
\end{equation}
where $a$ is the scale factor, $\rho$ is the density of the various
matter sources (radiation, dust) and $\rho_{\varLambda}$ is vacuum
energy density.

The constant $\rho_{\Lambda}$ causes accelerated expansion, but its
presence lacks physical ground. At the present epoch the matter density
$\rho$ is mainly dust (both baryonic and dark). The density of dust
dilutes as $a^{-3}$, therefore causes deceleration, which is physically
attributed to the attractive force of gravity between cosmic matter,
be it mostly dark. This view on cosmic matter (deceleration by attraction)
is supported by the well known Newtonian interpretation of the dust
term on the basis of the shell theorem when applied to an infinite
homogeneous isotropic universe, causing the field inside a spherical
cavity to be uniformly zero. Hence, a particle at the surface of an
arbitrary spherical section of the universe is effectively only attracted
by the matter interior this sphere and so is attracted to the center
of the sphere. Since this applies to arbitrary spherical sections,
cosmic matter pulls together by the force of gravity, according Newton. 

Though perhaps intuitive (matter attracts other matter), the Newtonian
view rests on the dubious assumption of instantaneous ``action at
a distance'' in an infinite universe. Instead, we consider propagation
of gravity at the speed of light in a universe of finite age, thus
assume a propagating gravitational horizon. However, in the presence
of a horizon, application of the shell theorem fails, therefore the
Newtonian derivation collapses (cf. \cite{Telkamp}). But actually,
this is what one would expect, since due to symmetry the gravitational
field in the perfectly homogeneous isotropic universe is zero everywhere,
which makes the concept of deceleration by gravitational attraction
questionable. The same argument applies to a ``repulsive force''
induced by dark energy. 

Zero field points at constancy of the kinetic energy of recession;
no energy is being exchanged. Within Newtonian context mass inertia
is invariant, so constant kinetic energy implies constant recession
velocities, i.e. a coasting universe, as pointed out earlier by Layzer
\cite{Layzer}. This obviously can not explain the \textgreek{L}CDM
model. 

In a Machian context however mass inertia is a relational property
which depends on the ``distribution of matter'', and so is likely
to evolve in an expanding universe. In the approach taken by Schrödinger
\cite{Schroedinger}, Machian kinetic energy is a relational and mutual
property between any pair of point particles ($m_{i}$,$m_{j}$) and
is defined 
\begin{equation}
T_{ij}\equiv\tfrac{1}{2}\mu_{ij}\dot{r}_{ij}^{2}\,,\label{eq:Tij-1-1}
\end{equation}
where $r_{ij}$ denotes the proper radial distance (separation) of
the particles. Crucial here is that from a relational point of view
\textit{only} radial motion is meaningful in the physical relationship
of the two point particles. This ontological notion is due to Bishop
Berkeley \cite{Berkeley,Telkamp}. Machian inertia $\mu_{ij}$ is
also a relational and mutual property between any two particles and
is defined 
\begin{equation}
\mu_{ij}\,\equiv\,m_{i}\frac{\varphi_{j}(r_{ij})}{\varphi_{e\!f\!f}}\,=\,m_{j}\frac{\varphi_{i}(r_{ij})}{\varphi_{e\!f\!f}}\,=\,\frac{-Gm_{i}m_{j}}{\varphi_{e\!f\!f}\,r_{ij}},\label{eq:muij}
\end{equation}
where the effective potential serves as a normalization parameter
which preserves consistency with Newtonian inertia \cite{Telkamp}.
Since only the radial component of motion contributes to the kinetic
energy between the two particles, the two perpendicular components
of motion do not contribute. This implies that in any peculiar motion
of a particle $m$ relative to the cosmic sphere effectively only
a fraction $\frac{1}{3}$ of the cosmic potential contributes to the
inertia of the particle relative to the universe. Hence, 
\begin{equation}
\varphi_{eff}={\textstyle \frac{1}{3}}\varphi,\label{eq:Phi eff}
\end{equation}
as explained in more detail in \cite{Telkamp}. 

Just by geometrical argument, the density of dust particles dilutes
as $a^{-3}$. This, however, only regards the number of particles,
but from a Machian perspective this is not necessarily true for the
\textit{energy} density of the dust; Eq.(\ref{eq:muij}) expresses
that mass inertia is a mutual property between particles and is proportional
to their mutual potential energy. This means inertial mass does not
exist without the other particle. This is the radical difference with
Newton's and Einstein's theory. It is a very strong notion, in light
of which it appears conceivable that the energy density of dust dilutes
at a rate different from $a^{-3}$. This applies equally to other
forms of matter, as we will argue. 

\noindent \bigskip{}

According \cite{Telkamp}, the Friedmann equation can be derived from
the Machian energy between a (unit) test mass $m$ and all the receding
cosmic matter within the gravitational horizon. Machian kinetic energy
between $m$ and all the cosmic matter within the horizon is \cite{Telkamp}
\begin{equation}
T={\textstyle \frac{3}{4}}\chi_{g}^{2}\dot{a}^{2},\label{eq:T}
\end{equation}
where $\chi_{g}$ is the comoving distance to the horizon (the symbol
$m$ is omitted for simplicity). The potential energy of $m$ due
to the cosmic masses within the horizon is according the Newtonian
definition, i.e., 
\begin{equation}
V=\varphi=-2\pi G\rho a^{2}\chi_{g}^{2}.\label{eq:V}
\end{equation}
In a strict Newtonian sense the matter density regards ordinary matter
only, but here we assume the density parameter $\rho$ includes all
different types of matter. The Machian energy equation is thus
\begin{equation}
T+V=\tfrac{3}{4}\chi_{g}^{2}\dot{a}^{2}-2\pi G\rho a^{2}\chi_{g}^{2}=E=const.,\label{T+V=00003DE-1}
\end{equation}
where we assume total energy $E$ is a conserved quantity. Similar
to the usual treatment of vacuum energy, we include total energy in
the extended density parameter, i.e. $\rho_{e}\!=\!\rho+\rho_{E},$
where 
\begin{equation}
\rho_{E}=E/2\pi Ga^{2}\chi_{g}^{2}\label{eq:rhoE}
\end{equation}
is the density of total energy. Associated with this density is the
potential 
\begin{equation}
\varphi_{E}=\!-2\pi G\rho_{E}a^{2}\chi_{g}^{2}.\label{eq:Phi E}
\end{equation}
Then the ``extended'' potential becomes 
\begin{equation}
\varphi_{e}=\varphi+\varphi_{E}=\!-2\pi G(\rho+\rho_{E})a^{2}\chi_{g}^{2}.\label{eq:Phi e}
\end{equation}
The Machian energy equation (\ref{T+V=00003DE-1}) can be rewritten
\begin{equation}
\tfrac{3}{4}\chi_{g}^{2}\dot{a}^{2}\,=\,2\pi G(\rho+\rho_{E})a^{2}\chi_{g}^{2},\label{eq:Ext energy}
\end{equation}
which is in fact the Friedmann equation 
\begin{equation}
\frac{\dot{a}^{2}}{a^{2}}={\textstyle \frac{8}{3}}\pi G(\rho+\rho_{E}).\label{eq:Fr eq}
\end{equation}
The intriguing aspect of Eq.(\ref{eq:rhoE}) is that the identity
of total energy density depends on the evolution of the horizon $\chi_{g}$.
For instance, if $\chi_{g}=const.$ then according Eq.(\ref{eq:rhoE})
we have $\rho_{E}\propto a^{-2}$, representing curvature energy,
like in the Newtonian case. In a de Sitter universe, however, the
proper distance to the event horizon is constant, i.e., $R_{g}\equiv a\chi_{g}=const.$,
therefore $\chi_{g}\propto a^{-1}$, which implies $\rho_{E}=const.$;
total energy density acts as constant vacuum energy density $\rho_{\Lambda}$,
i.e., as a cosmological constant. Hence, to determine the actual identity
of $\rho_{E}$ we must identify $\chi_{g}(a)$.

\medskip{}

\noindent In contrast with the Newtonian definition of the potential,
Sciama \cite{Sciama} derived, by a gravitational analog of Maxwell's
theory, a universally constant value of the cosmic potential equal
to 
\begin{equation}
\varphi_{u}=-c^{2}.\label{eq:phi u}
\end{equation}
It is therefore interesting to see how Sciama's constant potential
fits in, since constancy of the cosmic potential appears to be a fundamental
property of spacetime, just like the universal constancy of the speed
of light. The obvious question is thus how $\varphi_{u}$ relates
to the potential $\varphi_{e}$. 

Recalling that zero exchange of energy in the homogeneous isotropic
universe points at constancy of kinetic energy, the assumed conservation
of total energy indeed implies that also potential energy is constant.
We are thus strongly led to assume $\varphi_{e}$ is constant, i.e.,
\begin{equation}
\varphi_{e}=k\varphi_{u}=-kc^{2},\label{eq:phiu=00003Dkphie}
\end{equation}
where $k>0$ is a constant to be determined. Given Eq.(\ref{eq:Phi e})
and constancy of $E$, it follows that $\varphi=-2\pi G\rho a^{2}\chi_{g}^{2}$
is constant too. Hence, the immediate consequence of this is that
within the present Machian physics all types of matter-energy satisfy
a common relationship, 
\begin{equation}
\rho(a)\propto\rho_{E}(a)\propto a^{-2}\chi_{g}^{-2}(a).\label{eq:rho=00003DrhoE}
\end{equation}
This means densities dilute uniformly, at a fixed ratio. This may
seem wrong, as it goes totally against the established (as good as
undisputed) density-scale relations, like $\rho_{r}\propto a^{-4}$
for radiation and $\rho_{m}\propto a^{-3}$ for pressureless matter.
But recall that in Machian physics energy is \textit{not} an intrinsic
property of the particle, rather it is a property of its interactions
with all other particles of the universe. Therefore, it is leading
that the total energy associated with this cosmic interaction is constant,
instead of the un-Machian notion that isolated particles represent
energy (which in fact gives rise to presumed loss of photon energy
in the expanding universe, as well as creation of energy due to a
cosmological constant). If energy is about cosmic interaction, then
indeed the distance to the horizon is an essential parameter in the
definition of energies, as above.

\medskip{}

Anything beyond the horizon is causally disconnected from us, which
is the case if we assume the proper distance to the horizon propagates
away at the speed of light, i.e.,
\begin{equation}
\dot{R}_{g}=c.\label{eq:horizon}
\end{equation}
Since we consider gravity propagating at the speed of light, the horizon
$\chi_{g}$ behaves as a light front, so satisfies the condition of
a null geodesic. That is, given the FLRW metric $ds^{2}=c^{2}dt^{2}-a^{2}d\chi^{2}$,
the horizon propagates locally at the speed of light, i.e.,
\begin{equation}
a\dot{\chi}_{g}=\pm c,\label{nul-geodesic}
\end{equation}
where $+c$ holds for a particle horizon, while $-c$ applies to an
event horizon. From the Machian energy equation (\ref{eq:Ext energy})
and constancy of the extended potential, Eqs.(\ref{eq:Phi e},\ref{eq:phiu=00003Dkphie}),
we obtain the recession velocity ($>0$) of a galaxy at (and just
crossing) the horizon
\begin{equation}
\chi_{g}\dot{a}=\sqrt{{\textstyle \frac{4}{3}}k}\:c.\label{eq:Vrecession}
\end{equation}
The sum of Eq.(\ref{nul-geodesic}) and Eq.(\ref{eq:Vrecession})
is the speed of the proper distance to the horizon, i.e.,
\begin{equation}
\dot{R}_{g}=\chi_{g}\dot{a}+a\dot{\chi}_{g}=(\sqrt{{\textstyle \frac{4}{3}}k}\pm1)\:c.\label{eq:horizon-1}
\end{equation}
then, given Eq.(\ref{eq:horizon}), the only remaining choice of $k>0$
is 
\begin{equation}
k=3,\label{eq:k}
\end{equation}
while the sign of local speed of gravity must be negative. Thus recession
velocity of galaxies at the horizon Eq.(\ref{eq:Vrecession}) is twice
the speed of light

\begin{equation}
\chi_{g}\dot{a}=2c,\label{eq:Vrec=00003D2c}
\end{equation}
meaning that the horizon is at twice the Hubble distance. Since by
Eq.(\ref{eq:horizon-1}) the local speed of gravity must be negative
to meet $\dot{R}_{g}=c$, i.e.,
\begin{equation}
a\dot{\chi}_{g}=-c,\label{eq:Vinbound}
\end{equation}
we find that the horizon is an event horizon. This also follows from
the solution of Eqs.(\ref{eq:Vrec=00003D2c},\ref{eq:Vinbound}),
i.e.,
\begin{equation}
\chi_{g}(a)=\chi_{go}a^{q},\label{eq:Xg=00003DXgoaq}
\end{equation}
where $\chi_{go}$ is the present value of $\chi_{g}$ and where $q$
is the (constant) deceleration parameter. Substituting Eq.(\ref{eq:Xg=00003DXgoaq})
in Eq.(\ref{eq:Vinbound}) and substituting Eq.(\ref{eq:Vrec=00003D2c})
gives $-c=a\dot{\chi}_{g}=q\chi_{g}\dot{a}=q2c$, hence deceleration
is constant and equals,
\begin{equation}
q=-{\textstyle \frac{1}{2}}.\label{eq:q}
\end{equation}
Therefore, the unique solution of $\chi_{g}(a)$ is 
\begin{equation}
\chi_{g}(a)=a^{-\frac{1}{2}}.\label{eq:Xg(a)=00003Da-1}
\end{equation}
Note that $\chi_{g}(0)\!=\!\infty$, meaning that at Big Bang all
matter was causally connected and due to that we can still see last
images of sources which were once within the horizon (cf. Rindler
\cite{Rindler}), even though the horizon is at only twice the Hubble
distance. Substitution of Eq.(\ref{eq:Xg(a)=00003Da-1}) in Eq.(\ref{eq:rho=00003DrhoE})
yields, under the present Machian assumptions, the uniform identity
of all types of matter-energy,
\begin{equation}
\rho(a)\propto\rho_{E}(a)\propto a^{-1}.\label{eq:rho Mach}
\end{equation}
The simple cosmological model associated with this uniform matter-energy
is
\begin{equation}
H^{2}=H_{o}^{2}a^{-1},\label{eq:H}
\end{equation}
where $H=\dot{a}/a$ is the Hubble parameter and $H_{o}$ its present
value. 

\medskip{}

\noindent Thus we find that, subject to conservation of total Machian
energy and constancy of Machian kinetic energy:

i) Machian matter-energy of all types dilutes uniformly as $a^{-1}$,
thus gives rise to a constant accelerated expansion of the universe
at $q=-\frac{1}{2}$ from visible matter sources, i.e., without assuming
dark mater or dark energy. 

ii) In terms of the \textgreek{L}CDM model $q=-\frac{1}{2}$ means
$\varOmega_{m}=\frac{1}{3}$ and $\Omega_{\Lambda}=\frac{2}{3}$,
in good agreement with present observations (Planck data \cite{Ade}).
Nevertheless, the models deviate in the past and future because in
the \textgreek{L}CDM model $q$ varies with the scale factor. 

iii) Since $k=3$, it follows that the extended potential $\varphi_{e}=-3c^{2}$.
Therefore Sciama's potential $\varphi_{u}$ coincides, not surprisingly,
with the effective value of the extended potential, i.e.,
\begin{equation}
\varphi_{u}=\varphi_{e,eff}={\textstyle \frac{1}{3}}\varphi_{e}=-c^{2}.\label{eq:phie,eff}
\end{equation}

iv) The present Machian model of Eq.(\ref{eq:H}) has several favorable
properties. Total energy is conserved. It provides an explanation
of accelerated expansion without invoking dark matter or dark energy.
For the Machian model Eq.(\ref{eq:H}) it does not matter which fraction
of the extended potential $\varphi_{e}$ is due to dust (baryonic
or dark), photons or possible other components, which appear totally
interchangeable forms of energy. Furthermore, the model does not suffer
from the so called ``coincidence problem'', i.e., the unlikely coincidence
that $\Omega_{m}\thickapprox\Omega_{\Lambda}$ at present time (as
it in fact predicts $\Omega_{\Lambda}=2\Omega_{m}$ always). Neither
it needs an inflationary period to smooth out the universe after Big
Bang (the ``horizon problem''), since by Eq.(\ref{eq:Xg(a)=00003Da-1})
all matter of the universe was causally connected at Big Bang. Last
but not least, the model satisfies the Machian principle. 

\medskip{}


\begin{thebibliography}{1}
\bibitem{Sciama}D.W. Sciama. MNRAS, 113 , 34 (1953).

\bibitem{Rindler}W. Rindler, \textit{Relativity}. (Oxford University
Press 2006).

\bibitem{Berkeley}G. Berkeley (1721), \textit{De Motu}. English translation
by Arthur Aston Luce. (Nelson, London, 1951).

\bibitem{Schroedinger}E. Schroedinger. Die Erfüllbarkeit der Relativitätsforderung
in der klassischen Mechanik. Annalen der Physik, 382, 325 (1925);
For an English translation see: J. B. Barbour and H. Pfister, eds.,
Mach\textquoteright s Principle: From Newton\textquoteright s Bucket
to Quantum Gravity. (Birkenhauser, Boston, 1995).

\bibitem{Ade}P.A.R. Ade et al. (Planck Collaboration 2013 XVI), A\&A
(2015), arXiv:1303.5076v1.

\bibitem{Layzer}D. Layzer. Astronomical Journal 59, 268 (1954).

\bibitem{Telkamp}H.J.M. Telkamp (2016), arXiv:1604.01258.
\end{thebibliography}
\end{document}